# Energy Transfer Mechanism Under Incoherent Light Excitation in noisy Environments: Memory Effects in Efficiency Control

## Rajesh Dutta[1] and Biman Bagchi[2*]

**[1]Department of Chemistry, Purdue University, West Lafayette 47906, Indiana, USA.**

**[2]SSCU, Indian Institute of Science, Bangalore 560012, India.**

**\*Email: bbagchi@iisc.ac.in, profbiman@gmail.com**

## Abstract

Fluctuations in the energy gap and coupling constants in and between chromophores can play important role in the absorption and energy transfer across a collection of two level systems. In a noisy environment, fluctuations can control efficiency of energy transfer through several factors, including quantum coherence. Several recent studies have investigated the impact of light-induced stationary quantum coherence on the efficiency of transferring optical excitation to a designated "trap" state, crucial for subsequent reactions such as those in photosynthesis. However, these studies have typically employed either a Markovian, or a perturbative approximation for the environment induced fluctuations. In this study, we depart from these approaches to incorporate memory effects by using Kubo's quantum stochastic Liouville equation (QSLE). We introduce the effects of the decay of excitation (to the ground state) and the desired trapping that provides the direction of the motion of the excitation. In the presence of light-induced pumping, we establish a relation between the mean survival time, efficiency, and the correlation decay time of the bath-induced fluctuations. We find a decrease in the steady-state coherence during the transition from the non-Markovian regime (characterized by small values of fluctuation strength V and inverse of bath correlation time b) to the Markovian limit (where V and b are both large), resulting in a decrease in efficiency. Apart from V and b, the ratio $V^2/b$ plays a crucial role (as in Haken-Strobl model) in determining the mechanism of energy transfer and shaping the characteristics of the efficiency profile. We recover a connection between transfer flux and the imaginary part of coherences in both equilibrium and excited bath states, in both correlated and **uncorrelated bath models. We uncover a non-monotonic dependence of efficiency on site energy heterogeneity for both correlated and uncorrelated bath models.**

## I.     Introduction

That fluctuations in environment can play important role in quantum relaxation and quantum transport have been well-known for a long time.[1-5] When a system consisting of a large number of weakly coupled two level systems is placed under optical excitation (or pumping), both the absorption and subsequent transport of energy may depend crucially on the amplitude and time correlations of fluctuations in energy (the diagonal terms) and coupling coefficients. Recent developments in both theoretical and experimental studies have unveiled the crucial role played by the noisy protein environment within photosynthetic complexes, contributing to the enhancement of the excitation energy transfer process and maximizing overall energy transfer efficiency.[6-16]  Noise can induce resonance matching condition, thus fostering energy transfer. Thus, understanding the role of noise in our quest for conditions that allow efficient transfer of absorbed excitation energy from chromophores to reaction centre holds paramount importance. Earlier Haken, Strobl, Reinecker and Silbey developed elegant theories that included partly the effects of quantum motion in the diffusion of energy in a one-dimensional array of two-level systems.[1-3] A non-Markovian generalization of the same model was presented by Bagchi and Oxtoby.[17]

 Intense interest in this field is derived partly from the realization that the effects of noise may offer insights for advancing solar energy technologies such as photovoltaic devices and artificial photosynthesis. The efficientt energy transfer prompts an intriguing inquiry into identifying optimal conditions for maximizing the excitation energy transfer process within a multilevel extended quantum dissipative system.

The sub-picosecond time scale of quantum coherence observed in experiments on photosynthetic complexes[6-9] or conjugated polymers[18,19] raises several questions about the energy transfer process. One of these concerns is whether dynamical coherences induced by coherent ultrashort laser pulses themselves play a role in efficient energy transfer. Because, in natural conditions, energy transfer in photosynthetic light-harvesting occurs through incoherent sunlight. Furthermore, despite the short time scale of dynamical coherence, recent studies highlight the crucial role of steady-state coherences in non-equilibrium steady-state (NESS) for efficient energy transfer to the reaction center. The NESS envisages a steady flow of the excitation to the sink.

Over the past decade, there has been extensive research into the role of quantum coherence in excitation energy transfer, particularly in noisy environments. One of the significant challenges in studying excitation energy transfer lies in the non-perturbative treatment of intra-system or system-bath coupling, as the energy transfer occurs within the intermediate coupling regime for photosynthetic complexes. In our recent studies, we explored different coupling regime of energy transfer in photosynthetic system.[14,15] We observed that fluctuations not only have the capacity to disrupt coherence but, under specific circumstances, can also facilitate it. We demonstrated that temperature exerts the most significant influence in the intermediate coupling limit, where it can facilitate the transition from coherent to incoherent transfer. We also observed the decrease in temperature induces long-lasting quantum coherence, which subsequently triggers delocalization, resulting in the enhancement of coherence length. Coherence can be subdivided into two categories namely dynamical and static coherence. While dynamical coherence has been meticulously examined through the time evolution of the off-diagonal elements, it remains a transient phenomenon linked with the quantum beating observed in experiments. Due to the brief duration of quantum beating, the static coherence has a greater significance. The static coherence remains constant for long times and associates to stationary effects in equilibrium states or non-equilibrium steady states. We also established a connection between stationary coherences and non-canonical equilibrium population distributions.[20] Absence of light-induced energy transfer, results in equilibrium coherences in the long-time limit. The finite coherences at equilibrium are indicative of quantumness and defy replication in classical descriptions.[21,22] The finite coherences can give rise to non-canonical equilibrium populations.[23-25] However, in the presence of a source term or energy pumping, the energy transfer is predominantly governed by the NESS condition.

The inclusion of decay and trap possess several levels of theoretical challenge. First, the trapped state or reaction center must function as an energy sink, ensuring that no energy is regained once reaching this state. While this concept can be easily implemented in a classical framework via the rate description, its implementation becomes more intricate within a quantum formalism due to the Hamiltonian's hermiticity. Additionally, the complexity intensifies because of spatial and temporal correlation of the fluctuating environment in presence of light induced pumping. In presence of light induced pumping, the NESS has utmost importance in efficient energy transfer to the trap state.

Theoretical investigations in this domain can be broadly categorized into two groups. Firstly, the use of Markovian and perturbative equations of motion to explore optimal conditions for energy transfer in the absence of light-induced coherence. Gaab and Bardeen[26] predicted the optimal combination of trapping and dephasing for the minimal trapping time in N-site homogeneous chain systems by manipulating the trap position. In a series of insightful studies, Cao, Silbey, and co-workers explored the optimal conditions for energy transfer in both model and real photosynthetic complexes.[27-30] They studied linear and closed-loop configurations to observe the impact of phase modulation. Subsequently, they unveiled the role of the environment in the FMO complex using both seven and eight-site descriptions, considering dephasing models and incorporating spatial and temporal correlations to capture the real scenario in the energy transfer process within a dissipative environment. Cao et al. presented a scaling theory of average trapping time in weak and strong damping regimes, applying the theory to energy transfer in symmetric dendrimers. Aspuru-Guzik and their co-workers used a general dephasing model to investigate environment-assisted quantum transport in the FMO complex and binary trees.[31,32] Plenio and their co-workers explored the dependence of transfer efficiency on energy mismatch and dephasing rate for both linear and globally connected network models.[33,34] Chen and Silbey considered independent dichotomic noise, employing an exactly solvable approach to calculate population relaxation and determine the optimal conditions for energy transfer in symmetric dimers and photosynthetic complexes.[35] In our earlier study, we compared quantum and classical approaches for efficient energy transfer across various coupling limits, analysing both linear and cyclic models.[36]

On the other hand, there are studies that focus solely on investigating light-induced steady-state coherences, and its application towards efficient energy transfer in model and photosynthetic systems.[37-45] Zerah-Harush and Dubi revealed a universal origin for environment-assisted quantum transport in quantum networks with dephasing environment based on the connection between exciton current and occupation within a Markovian open quantum system framework.[41,42] They used a Lindblad equation and investigated the condition with and without static disorder. Cao and his colleagues examined the dimer model and its interaction with the environment, treating it as pure dephasing within the Haken-Strobl-Reineker framework.[43] They elucidated a coherence-flux-efficiency relationship and further deduced that the quantum coherence, as portrayed by the imaginary component of the density matrix, remains invariant across different basis sets for the dimer model. Jung and Brumer employed a hybrid methodology, merging the non-secular description of the Bloch–Redfield treatment with a Lindblad description for the remaining environment, in order to investigate the asymmetric donor-acceptor model.[44]

However, there are certain aspects in all these studies that necessitate further investigation. For example, most of these studies either rely on the Markovian master equation or employ a pure dephasing model approximation. However, in photosynthetic and conjugated polymer complexes, the perturbative and Markovian approximation may underestimate quantum coherence. Consequently, obtaining an accurate description of the optimal conditions for energy transfer may be challenging. Similarly, the pure dephasing model, though highly simplified, understates the complexity of energy transfer dynamics in photosynthetic complexes and conjugated polymers by neglecting spatial and temporal correlations. While non-Markovian aspects and higher-order system-bath correlations have been addressed using the Hierarchical equation of motion,[46,47] the study of the role NESS in obtaining optimal conditions for energy transfer remains unexplored.

In this study, our objective is to bridge the gap between these two types of investigations and examine the role of incoherent light-induced steady-state coherence in efficient energy transfer, employing the non-Markovian quantum stochastic Liouville equation. In this work, we consider correlated and uncorrelated bath models to explore the optimal conditions for energy transfer process. Utilizing a non-perturbative hierarchical equation of motion (HEOM) based on the quantum stochastic Liouville equation (QSLE),[48-52] we investigate various energy transfer regimes, including the strong coupling non-Markovian and weak coupling Markovian limits, in the presence of radiative decay and traps. Broad range of variation of parameters could be useful to study different type natural and artificial complexes.

In this work, we obtained several potentially intriguing results briefly outlined as follows

(1) In the case of a correlated bath, presence of excess coherence compared to an uncorrelated bath model leads to an increase in Mean Survival Time (MST) and reduction of efficiency. This distinctive behavior is not evident in the absence of incoherent light condition. The distinction becomes particularly pronounced at low intersite coupling (J), where the uncorrelated bath case exhibits an order of magnitude larger MST compared to the correlated bath model. This discrepancy arises from off-diagonal fluctuations playing a role similar to J, establishing connections between chromophores at small J, a scenario not achievable in the uncorrelated bath model.

(2) Different behavior with respect to the fluctuation strength (V) and rate of fluctuation is observed in the transition from non-Markovian (small V and b) to Markovian limit (large V and b). Efficiency profile experiences a decline when transitioning from the non-Markovian regime to the Markovian limit. In addition, the role of $V^2/b$ is revealed determining the nature of energy transfer.

(3) Transfer flux is defined with classical systems as a reference point, making it applicable for studying larger quantum networks. The transfer flux determines the pattern of efficiency with respect to the different coupling and relaxation parameters. Contribution of equilibrium and excited state coherences towards transfer flux can be disentangled for correlated bath model which is not possible for uncorrelated bath model.

The organization of the rest of the paper is as follows. In section II, we present the models and the theoretical formalism, including the Hamiltonian employed. In section III, we discuss the steady state population and coherences for correlated and uncorrelated bath model. In section IV, we elaborate the mean time and efficiency. In section V, we describe efficiency, transfer-flux and coherence relation. In section VI, we explain the effect of temperature. Section VI concludes with a discussion.

## II.    Model, Hamiltonian and equation of motion

For an isolated system driven by weak incoherent light one can write the Hamiltonian as follows

$$H(t) = H_S + H_{LM}(t) \qquad (1)$$

Where, the system Hamiltonian is given as

$$H_S = \sum_k \varepsilon_k \left|k\right\rangle\left\langle k\right| + \sum_{\substack{k,l \\ k\neq l}} J_{kl}\left|k\right\rangle\left\langle l\right| \tag{2}$$

Light-matter Hamiltonian can be written as

$$H_{LM}\left(t\right) = -\sum_k \boldsymbol{\mu}_k . \boldsymbol{E}_{LM}\left(t\right)\left[\left|0\right\rangle\left\langle k\right| + \left|k\right\rangle\left\langle 0\right|\right] \tag{3}$$

The transition dipole moment operator is given as

$$\boldsymbol{\mu}_{k0} = \boldsymbol{\mu}_{0k}^{\dagger} = \sum_k \boldsymbol{\mu}_k \left[\left|0\right\rangle\left\langle k\right| + \left|k\right\rangle\left\langle 0\right|\right] \tag{4}$$

The light matter Hamiltonian represents single excitation since we consider weak excitation.

In photosynthetic systems, the coherence time of the solar radiation is shorter than the time-scales of the system, coupling and the bath time-scale. In this limit, the incoherent light can be treated as the white-noise model for the description of the radiation as follows

$$\left\langle E_{LM}\left(t\right)\right\rangle = 0 \text{ and } \left\langle E_{LM}\left(t\right)E_{LM}^{*}\left(t'\right)\right\rangle = \frac{I}{\bar{\mu}^2}\delta\left(t-t'\right) \tag{5}$$

Here, I represents the pumping rate and $\bar{\mu} = \sqrt{\sum_k \mu_k^2}$ is magnitude of total dipole moment and the average is taken over the state of the incoherent light or the radiation density matrix.

Using perturbative approach and white noise approximation, one can obtain the equation of motion for isolated system as follows

$$\dot{\rho}\left(t\right) = -i\left[H_S,\rho\right] + \rho^0 \tag{6}$$

Throughout this work we assume $\hbar = 1$. Here, $\rho^0 = \frac{I}{\bar{\mu}^2}\boldsymbol{\mu}_{k0}\left|0\right\rangle\left\langle 0\right|\boldsymbol{\mu}_{0k}$ is pure coherent state

after single excitation by the incoherent light.

For an open quantum systems total Hamiltonian can be rewritten as follows:

$$H_{tot} = H_S + H_{LM}(t) + H_B + H_{\text{int}} \tag{7}$$

In the present study, we consider the total Hamiltonian in the interaction representation of the fluctuating bath Hamiltonian, so that the interaction $V$ is time dependent and can be modelled by a stochastic function with known statistical properties. In this interaction representation, the Hamiltonian can be represented as follows,

$$H(t) = H_{eff} + V(t) \tag{8}$$

where, $H_{eff} = H_S + H_{LM}(t)$

$$H_S = \sum_k E_k |k\rangle\langle k| + \sum_{\substack{k,l \\ k \neq l}} J_{kl} |k\rangle\langle l| - i\hbar \frac{k_t}{2} \left( |N\rangle\langle N| \right) - i\hbar \frac{k_d}{2} \sum_k |k\rangle\langle k| \tag{9}$$

where $E_k$ is the energy of an exciton localized at site $k$ and $J_{kl}$ is the intersite coupling between excitations at sites $k$ and $l$. N is the terminal site where trap is attached, $k_t$ and $k_d$ are the rate of trap and radiative decay respectively. For simplicity, we consider the decay rate is same for all sites or chromophore. In this study, we exclusively focus on the unidirectional energy transfer from the terminal chromophore to the trap through a trapping rate. We note that upon considering the interaction between the trap and the chromophore, significantly large trapping rate effectively disrupts the resonant condition essential for energy transfer.

The stochastic Hamiltonian V(t) is next decomposed into diagonal and off- diagonal fluctuations

$$V(t) = \sum_k |k\rangle\langle k| V_d(t) + \sum_{\substack{k,l \\ k \neq l}} |k\rangle\langle l| V_{od}(t) \tag{10}$$

The Hamiltonian is augmented with anti-Hermitian parts such as exciton recombination and exciton trapping to evaluate the mean time to reach the reaction center and energy transfer efficiency. We assume that fluctuations around each chromophore are spatially uncorrelated. In other words, the fluctuating elements in the Hamiltonian at each chromophore site are independent of one another. However, when considering temporal correlation, we examine two distinct scenarios: (1) In the correlated bath model, (Fig 1a) fluctuation around different chromophores are same or correlated all the times. (2) In the uncorrelated bath model, (Fig 1b) fluctuations are independent of each other around each chromophore.

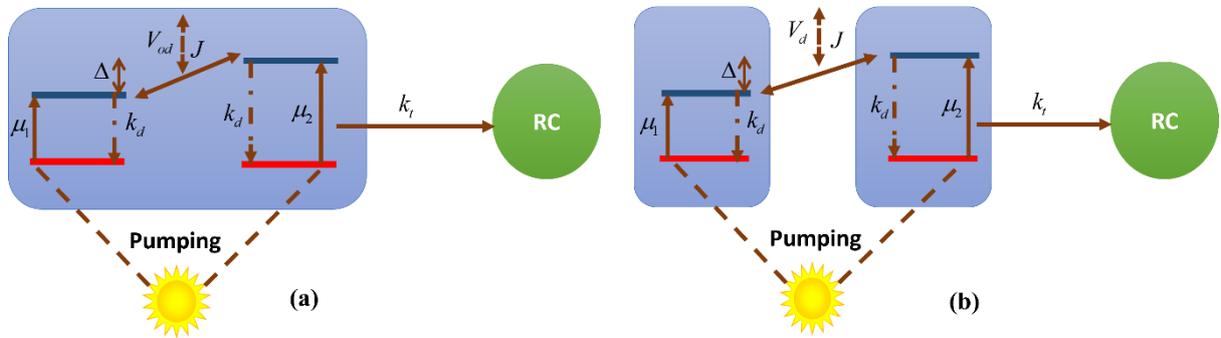

**Figure 1: Schematic picture of light driven efficient energy transfer to the reaction center in presence of fluctuating environment (a) correlated bath model (b) uncorrelated bath model. Here, each site or chromophore is modelled by a two-level system. Pale blue back ground signifies the fluctuating environment. Two sites in same environment designates correlated bath whereas two sites in different environment corresponds to the uncorrelated bath model. RC represents the reaction center or sink.**

In Fig. 1 the ground state of the molecule couples with excited state via transition dipole moment $\hat{\mu}_i$. Here, pumping rate $I$ has dimension of inverse of time. Multiplication of transition dipole moment and electric field provide the dimension of energy (Eq. 3).

We employ the non-perturbative quantum stochastic Liouville equation (QSLE) to treat quantum dissipative systems in different coupling regimes. We have previously discussed and employed the approach in detail in our earlier studies.[53,54] To maintain conciseness, we refrain from repeating the derivation but succinctly outline the fundamental steps leading to the final equations presented below (Eq. 11 and 12). First, we treat the light-matter interaction terms perturbatively, yielding a modified Quantum Liouville equation with the contribution of pure coherent states. The subsequent steps would be the same to the original stochastic Liouville equation, involving the joint probability distribution for the system and stochastic variable. This is followed by employing the Master equation of the stochastic variable to derive the ultimate reduced density matrix. The latter is then expanded in the eigenstates of the stochastic operator. These eigenfunctions are well-defined for two distinct processes: a Gaussian and a two-state Poisson. Remarkably, by considering only the first two eigenfunctions of the Gaussian Markov process, we deduce the equations for the Poisson bath. Here, we exclusively present the final equation of motion for correlated and uncorrelated bath model.

The coupled equations of motion (EOM) for dimer *with correlated fluctuation*, as follows

$$
\frac{d\sigma_m}{dt} = -i\left(E_1|1\rangle\langle1| + E_2|2\rangle\langle2|\right)^x \sigma_m - iJ\left(|1\rangle\langle2| + |2\rangle\langle1|\right)^x \sigma_m + \delta_{m,0}\rho^0
$$

$$
-k_t\left(|2\rangle\langle2|\right)\sigma_m - k_d\left(|1\rangle\langle1| + |2\rangle\langle2|\right)\sigma_m - iV_{od}\sum_{m'=0}^{1}(\delta_{m+1,m'} + \delta_{m-1,m'}) \times \left(\sum_{\substack{k,l \\ k\neq l}}|k\rangle\langle l|\right)^x \sigma_{m'} - mb\sigma_m
$$

(11)

where, we consider $\hbar = 1$, $O^x f = Of - fO$.

Similarly, one can derive the coupled equations of motion for dimer with uncorrelated fluctuation and is given as follows

$$
\frac{d\sigma_{jk}}{dt} = -i\left(E_1|1\rangle\langle1| + E_2|2\rangle\langle2|\right)^x \sigma_{jk} - iJ\left(|1\rangle\langle2| + |2\rangle\langle1|\right)^x \sigma_{jk} + \delta_{jk,00}\rho^0 - k_t|2\rangle\langle2|\sigma_{jk} - k_d\left(|1\rangle\langle1| + |2\rangle\langle2|\right)\sigma_{jk}
$$

$$
-iV_d\sum_{j'=0}^{1}\left(\delta_{j+1,j'} + \delta_{j-1,j'}\right)\left(|1\rangle\langle1|\right)^x \sigma_{j'k} - iV_d\sum_{k'=0}^{1}\left(\delta_{k+1,k'} + \delta_{k-1,k'}\right)\left(|2\rangle\langle2|\right)^x \sigma_{jk'} - jb_d\sigma_{jk} - kb_d\sigma_{jk}
$$

(12)

where, $V_{od}$, $V_d$ are the strengths of the fluctuation for correlated and uncorrelated bath fluctuation respectively and $b_{od}$ and $b_d$ corresponds to the rate of fluctuation. $\sigma$ is the reduced density matrix and suffix $m$ and $jk$ representing the eigenstate of the stochastic diffusion operator that can take 0 and 1 value for two level Poisson bath with eigenvalue 0 and $-b$. For a Gaussian stochastic process, the indices can vary from m=0 to infinity with eigenvalue -mb. We consider uncorrelated nearest neighbour fluctuations have same fluctuation strength and correlation time. The Kronecker delta terms in both correlated and uncorrelated bath case indicate the contribution of incoherent light induced pure coherent state.

## III. Steady state population and coherences:

As we are interested in NESS, population of the two sites for correlated bath model, can be written using Eq. (11) as follows

$$\sigma_0^{11} = \frac{\hat{\mu}_1^2 I}{k_1} - \frac{2J}{k_1} \operatorname{Im} \sigma_0^{12} - \frac{2V_{od}}{k_1} \operatorname{Im} \sigma_1^{12}$$
$$\sigma_0^{22} = \frac{\hat{\mu}_2^2 I}{k_2} + \frac{2J}{k_2} \operatorname{Im} \sigma_0^{12} + \frac{2V_{od}}{k_2} \operatorname{Im} \sigma_1^{12}$$

(13)

where, $k_1 = k_d$ and $k_2 = k_d + k_t$

One can easily prove from the above equations

$$I = k_1 \sigma_0^{11} + k_2 \sigma_0^{22}$$

(14)

Eq. (14) suggests the exciton pumping flux is equivalent to the exciton depletion flux.

Equation for the steady state coherences can be written as

$$\left(-i\Delta + \bar{k}\right)\sigma_0^{12} - iJ\left(\sigma_0^{11} - \sigma_0^{22}\right) - iV_{od}\left(\sigma_1^{11} - \sigma_1^{22}\right) = \hat{\mu}_1 . \hat{\mu}_2 I$$
$$\left(-i\Delta + \bar{k} + b_{od}\right)\sigma_1^{12} - iJ\left(\sigma_1^{11} - \sigma_1^{22}\right) - iV_{od}\left(\sigma_0^{11} - \sigma_0^{22}\right) = 0$$

(15)

Population difference in excited bath state in the above expression can be expressed as

$$\sigma_1^{11} - \sigma_1^{22} = -\left(\frac{2}{k_1 + b_{od}} + \frac{2}{k_2 + b_{od}}\right)\left[J \operatorname{Im}\sigma_1^{12} + V_{od} \operatorname{Im}\sigma_0^{12}\right] \tag{16}$$

Solution of the Eq. (15) can be obtained by using Eq. (13) and (16). The imaginary part of the steady state coherences in equilibrium and excited bath states are expressed as

$$\operatorname{Im}\sigma_0^{12} = \frac{QR - SN}{MQ - PN}$$

$$\operatorname{Im}\sigma_1^{12} = \frac{MS - PR}{MQ - PN} \tag{17}$$

where,

$$M = \left[1 + \frac{2J^2\bar{k}}{\bar{k}^2 + \Delta^2}\frac{k_1 + k_2}{k_1 k_2} + \frac{2V_{od}^2\bar{k}}{\bar{k}^2 + \Delta^2}\frac{k_1 + k_2}{k_1 k_2}\frac{(k_1 + k_2 + 2b_{od})}{(k_1 + b_{od})(k_2 + b_{od})}\right]$$

$$N = \frac{2JV_{od}\bar{k}}{\bar{k}^2 + \Delta^2}\left[\frac{(k_1 + k_2 + 2b_{od})}{(k_1 + b_{od})(k_2 + b_{od})} + \frac{k_1 + k_2}{k_1 k_2}\right]$$

$$P = \frac{2J(\bar{k} + b_{od})V_{od}}{(\bar{k} + b_{od})^2 + \Delta^2}\left[\frac{(k_1 + k_2 + 2b_{od})}{(k_1 + b_{od})(k_2 + b_{od})} + \frac{k_1 + k_2}{k_1 k_2}\right]$$

$$Q = \left[1 + \frac{2J^2(\bar{k} + b_{od})}{(\bar{k} + b_{od})^2 + \Delta^2}\frac{(k_1 + k_2 + 2b_{od})}{(k_1 + b_{od})(k_2 + b_{od})} + \frac{2V_{od}^2(\bar{k} + b_{od})}{(\bar{k} + b_{od})^2 + \Delta^2}\frac{k_1 + k_2}{k_1 k_2}\right]$$

$$R = \frac{\hat{\mu}_1 \cdot \hat{\mu}_2 I \Delta}{\bar{k}^2 + \Delta^2} + \frac{J\bar{k}}{\bar{k}^2 + \Delta^2}\frac{(\hat{\mu}_1^2 k_2 - \hat{\mu}_2^2 k_1)I}{k_1 k_2}$$

$$S = \frac{V_{od}(\bar{k} + b_{od})}{(\bar{k} + b_{od})^2 + \Delta^2}\frac{(\hat{\mu}_1^2 k_2 - \hat{\mu}_2^2 k_1)I}{k_1 k_2}$$

$$\tag{18}$$

and, $\bar{k} = \frac{(k_1 + k_2)}{2}$

Now, substitution of the imaginary part of the coherences in steady state population one can obtain the following expression.

$$\sigma_0^{11} = \frac{\hat{\mu}_1^2 I}{k_1} - \frac{2J}{k_1} \frac{QR - SN}{MQ - PN} - \frac{2V_{od}}{k_1} \frac{MS - PR}{MQ - PN}$$

$$\sigma_0^{22} = \frac{\hat{\mu}_2^2 I}{k_2} + \frac{2J}{k_2} \frac{QR - SN}{MQ - PN} + \frac{2V_{od}}{k_2} \frac{MS - PR}{MQ - PN}$$

(19)

For an uncorrelated bath model, the steady state-population and coherence relation can be written as follows

$$\sigma_{00}^{11} = \frac{\hat{\mu}_1^2 I}{k_1} - \frac{2J}{k_1} \operatorname{Im} \sigma_{00}^{12}$$

$$\sigma_{00}^{22} = \frac{\hat{\mu}_2^2 I}{k_2} + \frac{2J}{k_2} \operatorname{Im} \sigma_{00}^{12}$$

(20)

In this case, the coherence in excited bath states $\sigma_{mn}^{12}$ (m and n could be either 0 or 1) is coupled to the coherence in equilibrium bath states and thus cannot be separated. For both correlated and uncorrelated bath model, the first term (Eq. 13 and 20) indicates the contribution from the excitation and depletion at the same site or chromophore. However, the remaining terms are identical for both sites, with opposite signs indicating coherent transfer contributions.

## IV.    Mean survival time and efficiency:

Mean survival time require to reach the trap state can be defined using the steady state population as follows[42]

$$\langle \tau \rangle = \frac{1}{I} \sum_i \sigma_0^{ii}$$

(21)

Eq. (21) is either can be obtained from the stationary solution to the equation of motion or terms of residence time or integrated population. In general, the meantime can be obtained by solving linear system of equation using Eq. (11) and (12) $\dot{\sigma} = L\sigma + \rho^0$ with the assumption of steady state i.e. $\dot{\sigma} = 0$. As a result, one can obtain the stationary solution as $\sigma = L^{-1}\rho^0$. In our earlier study, the mean survival time is defined for local molecular excitation.[37] However, in this work the excitation is induced by stationary sunlight where $\rho^0$ consists of delocalized excitations as well as interference between excitations.

Efficiency or quantum yield can be defined as follows[42]

$$Q = \frac{k_t \sigma_0^{22}}{k_d \sigma_0^{11} + \left(k_d + k_t\right)\sigma_0^{22}} \tag{22}$$

Here, efficiency is defined in terms of probability of trapping. We note that the denominator of Eq. (22) represents the total depletion flux, incorporating both decay and trapping effects, thereby equating to the pumping flux as expressed in Eq. (14). Therefore, the equation of efficiency or quantum yield simplifies to

$$Q = \frac{1}{I} k_t \sigma_0^{22} \tag{23}$$

In case of uncorrelated bath model, the mean time and efficiency equation can be replaced by $\sigma_{00}^{ii}$ and $\sigma_{00}^{22}$.

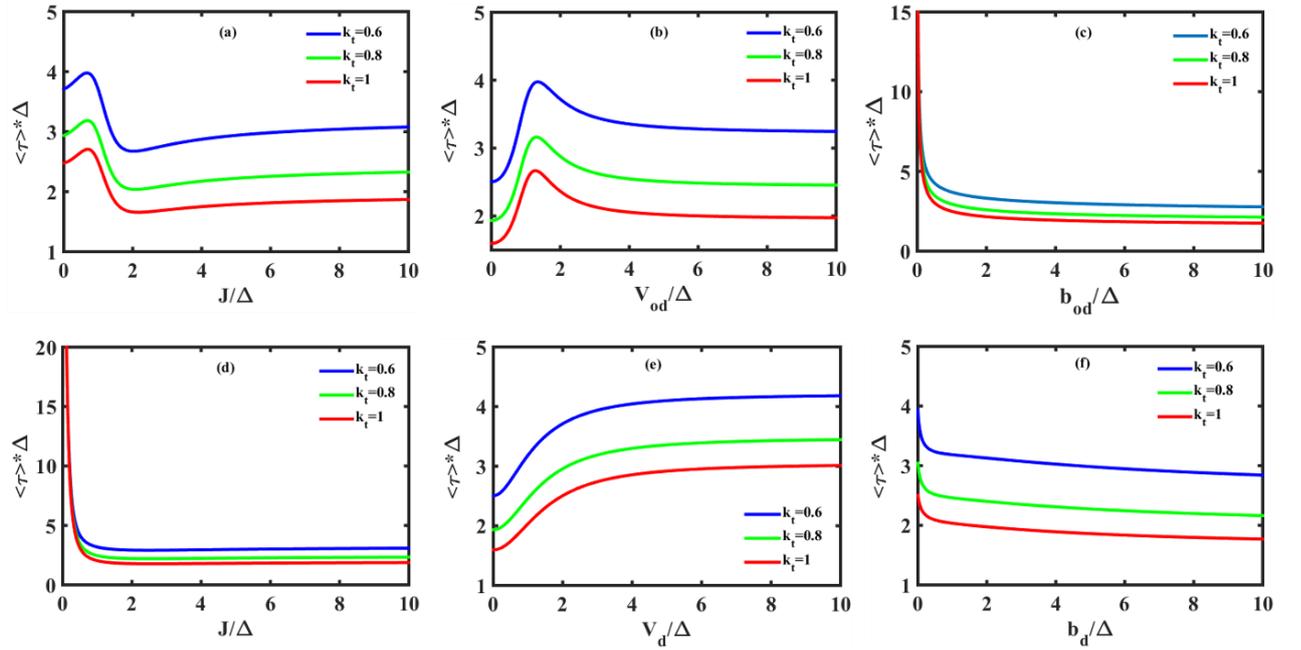

**Figure 2: Panel (a)-(c) and (d)-(f) show plot of mean survival time at different trapping rate and decay rate $k_d$=0.01 for correlated and uncorrelated bath model. (a) mean survival time vs. inter-site coupling at fluctuation strength $V_{od}$=1 and bath relaxation rate, $b_{od}$=1, (b) mean survival time vs. fluctuation strength ($V_{od}$) at J=1 and $b_{od}$=1, (c) mean survival time vs. bath relaxation rate ($b_{od}$) at J=1 and $V_{od}$=1. (d) mean survival time vs. inter-site coupling at fluctuation strength $V_d$=1 and bath relaxation rate, $b_d$=1 (e) mean survival time vs. fluctuation strength ($V_d$) at J=1 and $b_d$=1, (f) mean survival time vs. bath relaxation rate ($b_d$) at J=1 and $V_d$=1.**

In the case of photosynthetic energy transfer from the chlorophyll antennas to the reaction center, both the rate of trapping and energy relaxation is generally of the order of ~ 1ps$^{-1}$. However, the recombination rate or relaxation rate towards ground state could be lower than 1ps$^{-1}$ by several order of magnitude, and the associated rate is ~ 1ns$^{-1}$. We picked the time scales of trapping and decay such that the difference in time scale is similar to those in photosynthetic energy transfer.

Here, we chose the parameter space in dimensionless units, where energy parameters are scaled by the energy gap parameter $\Delta$, which is assumed to be unity all through. Thus, the coupling parameter $J$ and stochastic parameters $V_d$ and $V_{od}$ are all scaled by $\Delta$=1. We have approached the Markovian limit by increasing the rate of fluctuations, denoted by $b_d$ (diagonal, that is, site energy) and $b_{od}$ (off-diagonal, intersite coupling) parameters and strength of fluctuation $V_d$ and $V_{od}$. The theoretical results are discussed below.

In Figure **2(a)-(c)**, the mean survival time (MST) is plotted with respect to intersite coupling, fluctuation strength, and the rate of fluctuation for the correlated bath model. Investigations into the photosynthetic system reveal that energy transfer occurs in the coupling regime where the intersite coupling and system-bath coupling strength are similar.

In Figure **2(a)**, an initial rise in MST is noted with the increase in J at intermediate values of V and b. This suggests that the excitation at both sites fails to reach the trap state efficiently due to the activation of intersite coupling, leading to quantum interference between pathways. However, as intersite coupling increases, the competition between V and b with J minimizes the MST. To confirm the role of quantum coherence, the MST is also evaluated at large V and b limit, where the destruction of quantum coherence eliminates the initial rise in mean survival time. In Figure 2(b), a similar feature is observed in MST vs. J as V is directly associated with J. At low values (close to J) of off-diagonal fluctuation, MST increases. Further increases in V result in a decrease in MST. The ratio $V^2/b$ also determines the nature of transport. At small J and large b, the plot exhibits a different pattern, with MST decreasing with increasing V and saturating after $V^2/b > 2J$, indicating a transition from coherent to incoherent transport. For Figure 2(c), at the intermediate coupling regime (J and V equal), an increase in b leads to a decrease in MST, with a slow decrease observed at large values of b. In the low J and large V limit, mean survival time decreases with increasing b until the ratio $V^2/b$ is close to 2J. Beyond this point, further increases in b result in an increase in MST. Additionally, the dependence of MST on oscillator strength is verified, indicating that a large oscillator strength at the terminal site decreases MST.

The MST plot for an uncorrelated bath is illustrated in Figures 2(d)-2(f). In both correlated and uncorrelated bath models, an increase in trapping rate results in a reduction in MST. In Figure 2(d), as the coupling constant J increases, MST experiences an initial decrease, reaching a saturation point when J equals both V and b. This saturation indicates the existence of an intermediate coupling regime. A subsequent increase in b while keeping V constant leads to a further diminishing of MST. Emphasizing the earlier observation in the correlated bath case, the ratio $V^2/b$ plays a pivotal role in determining the nature of transport. Notably, the initial hump observed in Figures 2(a) and 2(b) is absent in both Figures 2(d) and 2(e), attributed to the protectionon of coherence due to the bath correlation. In Figure 2(e), an increase in V corresponds to a increase of the mean time, indicative of the destruction of coherence. This observation is further clarified by examining Figure 2(f), where an increase in b results in a reduction of MST, attributed to the destruction of fluctuations. Intriguingly, at large values of V with fixed b, MST exhibits a subsequent increase.

In order to further investigate the competition between J, V and b, plot of quantum yield or efficiency is shown in three dimension. Interestingly, energy transfer in photosynthesis is an efficient process because of large time scale separation between trapping and recombination.[55] Hence, to understand the competition between parameters clearly, comparable trapping and recombination or decay rates are assumed.

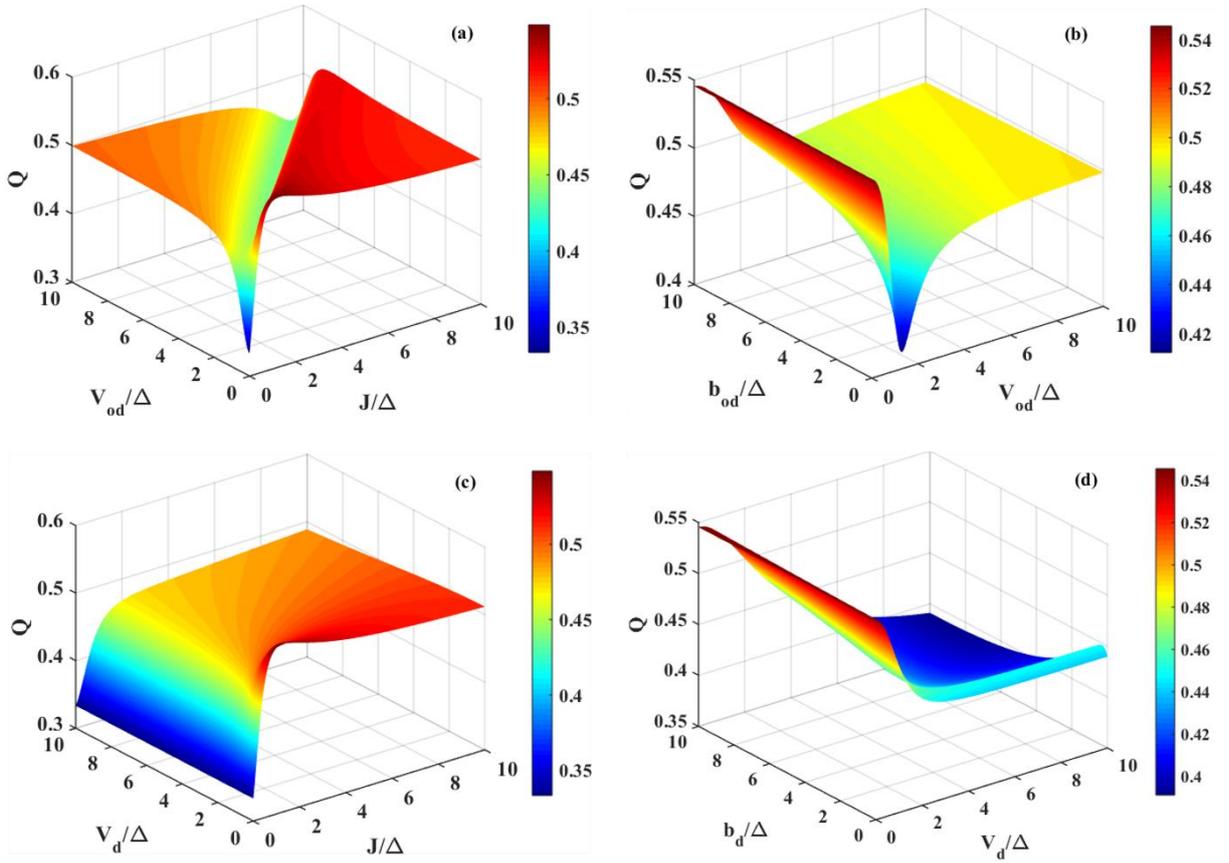

**Figure 3: Panel (a)-(b) show three-dimensional plot of quantum efficiency (Q) for correlated bath model at trapping rate $k_t$=1 and decay rate $k_d$=0.5. (a) quantum efficiency (Q) vs. intersite coupling (J) vs. fluctuation strength ($V_{od}$) at $b_{od}$=1 (b) quantum efficiency (Q) vs. fluctuation strength ($V_{od}$) vs. bath relaxation rate ($b_{od}$) at J=1. Panel (c)-(d) show three-dimensional plot of quantum efficiency (Q) for correlated bath model at trapping rate $k_t$=1 and decay rate $k_d$=0.5. (c)**

**quantum efficiency (Q) vs. intersite coupling (J) vs. fluctuation strength ($V_d$) at $b_d$=1 (d) quantum efficiency (Q) vs. fluctuation strength ($V_d$) vs. bath relaxation rate ($b_d$) at J=1.**

The efficiency exhibits an opposing trend to that of MST. In Fig. 3(a), the efficiency of the correlated bath model is plotted against intersite coupling (J) and fluctuation strength (V). Figure 3(a) clearly illustrates that, for a fixed b and V, efficiency experiences an initial dip followed by an increase, eventually reaching saturation at large J. This initial dip is attributable to the onset of J, giving rise to quantum interference between pathways. Upon increasing V, both the minima and maxima shift towards higher J values due to the ongoing competition between quantum interference and environmental fluctuation. Conversely, at a fixed V and J, an increase in b results in the disappearance of the initial minima, followed by an overall enhancement in efficiency. This improvement is due to the destruction of quantum coherence induced by the rapid bath fluctuations. Similarly, according to Fig. 3(b), there is a decrease in efficiency when V increases for a fixed J and b. However, as b increases, efficiency shows an upward trend. Nonetheless, an increase in J leads to a further decline in efficiency.

In the case of Fig. 3(c), efficiency shows an enhancement with the increase in J, followed by a saturation at large J for a fixed V and b. Similar to that of the correlated bath model, with the increase in V, efficiency decreases. Contrary to this, an increase in b shows an overall increase in efficiency. Analyzing Fig. 3(d), it can be inferred that with the increase in V at fixed b and J, efficiency decreases gradually and saturates at large V. In contrast, with the increase in b at fixed J, saturation occurs at a higher value of V and lower efficiency due to the decrease in the $V^2/b$ ratio. At the limit V and b$\rightarrow\infty$, our model reaches the Markovian limit. In this scenario, rather than considering the individual parameters V and b, the dynamics of the fluctuating environment can effectively be described by a single parameter, as demonstrated in the Haken-Strobl-Reineker-Silbey model.[1-3]

In the non-Markovian limit (characterized by low values of V and b), as depicted in Fig. 3(a) and 3(c), the efficiency is notably higher compared to the Markovian limit (where V and b are large), as illustrated in Fig. 3(b) and 3(d). According to Eq. (13), (20), and (23), the increase in efficiency is associated with an increase in the imaginary part of the steady-state coherences. Based on previous studies,[14,15] we know that the memory effect enhances dynamical coherence. Therefore, we can infer that the memory effect promotes both dynamical and steady-state coherences.

We also explore the impact of asymmetry or site energy heterogeneity (Δ) on efficient energy transfer. In both the correlated and uncorrelated bath models, efficiency exhibits a non-monotonic relationship with Δ. Efficiency increases with increasing Δ, reaching a maximum at an intermediate value close to J, then decreases with further increases in Δ. Interestingly, for a fixed value of J and V, an increase in b leads to enhanced efficiency due to the decrease in the $V^2/b$ ratio.

## V.    Transfer flux and multi-chromophoric system:

Transfer flux plays essential role in characterizing the efficient energy transfer in extended systems or molecular network. Cao, Silbey and co-workers defined transfer flux[23] (F) for classical system using integrated classical population as follows

$$F_{ij} = k_{ij}\tau_j - k_{ji}\tau_i \tag{24}$$

Where, k is the rate of transfer between different sites and $\tau$ is the integrated population or residence time. The equivalent transfer flux can be obtained for quantum system of correlated bath model as follows

$$F_{ij} = 2J \operatorname{Im} \sigma_0^{ij} + 2V \operatorname{Im} \sigma_1^{ij} \tag{25}$$

In the case of correlated bath model, both the coherences in equilibrium and excited bath states play a crucial role.

For uncorrelated bath the transfer flux can be simply written as

$$F_{ij} = 2J \, \mathrm{Im} \, \sigma_{00}^{ij} \qquad \qquad (26)$$

In the case of uncorrelated bath model, the coherences in excited bath state indirectly contributes towards the transfer flux via the coupling between element of density matrix. In the absence of source or pumping term, the transfer flux is zero for our QSLE dynamics at steady state because of detailed balance condition. Excitation energy transfer in photosynthetic complex is governed by NESS which is driven by absorbed photons.

Now, using the definition of quantum yield or efficiency one can obtain the expression of efficiency in terms of transfer flux as follows

$$Q = k_t \left[ \frac{\hat{\mu}_2^2}{k_2} + \frac{F_{12}}{I k_2} \right] \qquad \qquad (27)$$

The expression of efficiency can be easily extended to the linear chain of multi-chromophoric systems where the trap site is attached to the terminal site (N).

$$Q_{multi} = k_t \left[ \frac{\hat{\mu}_N^2}{k_N} + \sum_{i \neq j} \frac{F_{ij}}{I k_N} \right] \qquad \qquad (28)$$

The two terms in Eq. (28) can be entitled as monomer and transfer contribution respectively. The plot of transfer contribution with respect to the intersite coupling, fluctuation strength or rate of fluctuation will be similar in nature to that of efficiency.

## VI. Effect of Temperature

Temperature effect in the equation motion can be included through different parameters such as J and V as well as temperature corrected equation of motion. While J exhibits weak dependency on temperature, a decrease in temperature leads to molecules being packed more closely together. Consequently, the overlap between wave functions increases, resulting in an enhancement of J. However, V has a strong temperature dependence and decreases rapidly with as the temperature decreases. Therefore, a decrease in temperature essentially indicates an increase in coherence as well as transfer flux.

On the other hand, the stochastic theory deals with the fluctuation of energy states but does not incorporate dissipation terms in the equation of motion. Dissipation arises alongside fluctuation due to the system's interaction with a heat bath. In equilibrium, energy balance is achieved through fluctuation-dissipation dynamics. However, the dissipation term is absent in the stochastic Liouville equation as it neglects back reaction from the system to the bath. Consequently, at finite temperatures, the system fails to reach equilibrium in the long-time limit. However, this discrepancy can be eliminated considering bath as a collection of Harmonic oscillators. In this case, the bath correlation function has both real and imaginary part leading to the symmetric and antisymmetric part of the correlation function. The antisymmetric part of the bath correlation function is responsible for the dissipation. High temperature approximation results to the exponential decay of symmetric and antisymmetric part of the correlation. Temperature corrected equation of motion was derived by Taniumra and Kubo[56] and can be written as follows

$$\frac{d\sigma_m}{dt} = -iH_S^x \sigma_m - iV_d \left( \sum_k |k\rangle\langle k| \right)^x \sigma_{m+1} - imV_d \left[ \left( \sum_k |k\rangle\langle k| \right)^x - \frac{i\beta b}{2} \left( \sum_k |k\rangle\langle k| \right)^o \right] \sigma_{m-1} - mb\sigma_m$$

(29)

where, $O^x f = Of - fO$ and $O^o f = Of + fO$. If one neglects the imaginary term that comes from antisymmetric part of the bath correlation the equation becomes equivalent the QSLE for Gaussian bath states. Eq. (29) can be extended the to incorporate the decay, trap and incoherent light induced contributions. Eq. (29) accurately determines energy transfer dynamics at room temperature for photosynthetic systems.[7] QSLE overestimates the coherence with respect to the temperature corrected equation of motion which will essentially changes the MST and efficiency profile. In our earlier study, we showed the temperature correction equation leads to the non-canonical thermal distribution due to the system-bath correlation.[16] As a result, the steady-state coherence arises from the combined effects of thermalization and light-induced NESS conditions in the presence of trapping and decay processes.

## VII.  Conclusions:

Understanding the efficiency of energy transfer to the trap in the photosynthetic reaction system or its biomimetic analogue requires study of quantum transport in the presence of quantum coherence and energy decay. In the systems of interest, the study additionally has to deal with noisy environment that gives fluctuations in the diagonal and off-diagonal terms in our Hamiltonian. Upton a point, diagonal fluctuations can help in the absorption of light. However, fluctuations can lead to reduction in quantum coherence which can play a positive role in the transport to the trap. This is where correlations in and among fluctuations could play important role.

In this work, we employ Kubo's quantum stochastic Liouville equation (QSLE) to explore the light induced excitation energy transfer. Our objective of the study is the impact of different coupling limit in Markovian and non-Markovian limit, particularly in the presence of incoherent light-induced coherence for correlated and uncorrelated bath model. The motivation is to unravel the factors contributing to the high efficiency of the transfer to the reaction center in photosynthetic systems. Despite the apparent simplicity of the system, it reveals intricate dynamics. The key findings of this study are briefly outlined as follows

(1) In case of correlated bath model, MST shows an initial enhancement due to the additional quantum coherence induced by incoherent light at intermediate J, which was absent in excitation transfer in initially localized condition. Interestingly, MST with respect to the V shows different behavior in different regime. At the intermediate value of J and b, increase of MST was noted with the increase in fluctuation strength followed by a saturation at large V. Contrary to this behavior, at small J and large b, MST decreases with the increase in fluctuation strength till $V^2/b \approx 2J$ limit is attained and reaches a constant value at large V. The change of nature of MST indicates the coherent to incoherent transition going form strong coupling non-Markovian to weak coupling Markovian limit (large V and b).

(2) In case of uncorrelated bath model, a behavior similar to correlated bath can be observed from different MST plot. However, uncorrelated bath can diminish the coherence effectively than that of correlated bath and as a result the initial increase in MST is absent. At low value of J, larger value of MST was observed compared to correlated bath case. This is because, in case of correlated bath case at low J, the off-diagonal fluctuation strength establishes the connection between the sites and helps the energy transfer which cannot be possible by the diagonal fluctuation in case of uncorrelated bath.

(3) In both correlated and uncorrelated bath cases, the substantial oscillator strength near the trap state results in a decrease in mean survival time (MST), thereby facilitating efficient energy transfer. We also note a non-monotonic relationship between efficiency and site energy heterogeneity, with the maximum efficiency occurring near the intersite coupling.

(4) Efficiency or quantum yield exhibits the opposite trend to that of MST for both correlated and uncorrelated bath case. With the parameter value of decay and trapping rates related to the photosynthesis high efficiency close to unity and minute variation in the efficiency profile was observed. Efficiency shows dependence on the ratio $V^2/b$ in addition to the V and b separately. Memory protects steady-state coherence, resulting in an increase in efficiency.

(5) In case of correlated bath role of steady state coherence in equilibrium and excited bath state can be separated. However, in case of uncorrelated bath case, the effect of coherence in excited bath states towards the transfer flux is hidden into the contribution by the coherence in equilibrium bath states and as a result cannot be disentangled.

Finally, we conclude that we explored the effect of incoherent light induced NESS across different coupling limit ranging from strong coupling non-Markovian to weak coupling Markovian limit. Our investigation has unveiled numerous intriguing conditions for controlling transfer flux and energy transfer efficiency, offering potential insights for designing optimal artificial energy transfer networks. Especially, our study can effectively describe the trapping of excitation energy in the presence of spatial and temporal correlation. This work could be extended by the incorporation of temperature correction which will make the study closer to the complex biological networks at physiological conditions.

## ACKNOWLEDGEMENTS

BB thanks SERB (DST), India, for an Indian National Science Chair Professorship.